# Method of accelerator search for dark matter

Victor Kryshkin

*Institute for high energy physics, Protvino, 142281, Russia*

**Abstract**

A method to search for dark matter by studying the interaction of accelerator beam particles with residual gas in dependence on gas pressure is proposed. The sensitivity of the method is estimated.

## 1. Introduction.

There is a strong indication that a substantial part of the Universe exists in the form of dark matter. Notwithstanding the tremendous efforts to gain an understanding of the problem so far there is no conclusive proof of it. Up to now some alternative explanations (such as modified gravity) are still considered. So the main point in this study is to unambiguously prove that the observed effects can be explained by dark matter.

We propose to use particles of the dark matter as a target for accelerated particles (electrons or protons). Studying the loss of the beam particles due to interactions with residual gas in accelerator chamber in dependence on gas pressure and interpolating it to zero pressure one can estimate the contribution of some "invisible" stuff to the life time of the beam.

## 2. Evaluation of the method potential.

The dependence of beam particle loss due to the residual gas in accelerator chamber can be expressed in the following form:

$$N = N_0\,(1 - e^{l/\lambda}), \qquad (1)$$

where $N_0$ is the number of accelerated particles, $l$ is the length of the particle path, $\lambda$ is the nuclear interaction length of the residual gas in the accelerator chamber at the given pressure.

For air at normal conditions interaction length is $\lambda$ = 90 g/cm$^2$, density is $\rho$ = 0.129 *g*/*l* and $\lambda$=698 m. For arbitrary pressure the effective interaction length will be varying as

$$\lambda/P. \qquad (2)$$

At pressure $P = 10^{-5}$ Pa the effective nuclear length is $7\times10^{12}$ m. During one minute high energy particles will fly a path

$$L=3\times10^{10}\ \text{cm/s} \times 60\ \text{s}=1.8\times10^{10}\ \text{m}.$$

Because the path length $L$ is notably less than the nuclear path length of the residual gas in accelerator chamber the expression (1) can be rewritten as:

$$N=N_0\, L/(\lambda/P)=(N_0L/\lambda)P. \qquad (3)$$

For beam intensity $N_0 = 10^{14}$ the expression (3) will be

$$N = 2.6\times10^{21}\, P.$$

Fig. 1 shows the dependence of $N$ on $P$. In order to estimate the potential capability of the method let the precision of the particle current measurement during 1 minute is about 1% , then the final precision of the current measurements during a day (1440 repeated measurements) will be $2.6\times10^{-4}$. If there is exist beside a residual gas some additional source of particle absorption (like the dark matter) then the expression (3) will be presented in the form:

$$N= (N_0L/\lambda)P + (N_0L/\lambda_d), \qquad (4)$$

where $\lambda_d$ is the nuclear absorption length of the dark matter. The second term in the right part of the expression (4) does not depend on the residual gas pressure. Suppose the gas pressure has three setting ($10^{-5}$, $5\times10^{-5}$ и $10^{-4}$ Pa) and is measured with the 1% precision. Then the error in measurements of the second term in the expression (4) will be about $10^{-3}$. From there the minimal nuclear interaction length of the dark matter will be $1.8\times10^{27}$ m. From the ration

$$\lambda_d/\lambda_0=1.8\times10^{27}\ \text{m}/698\ \text{m}=2.6\times10^{24}$$

the minimal cross section that can be extracted by this method with quoted precision of the pressure and the beam current measurements is $10^{-44}$ cm$^{-2}$.

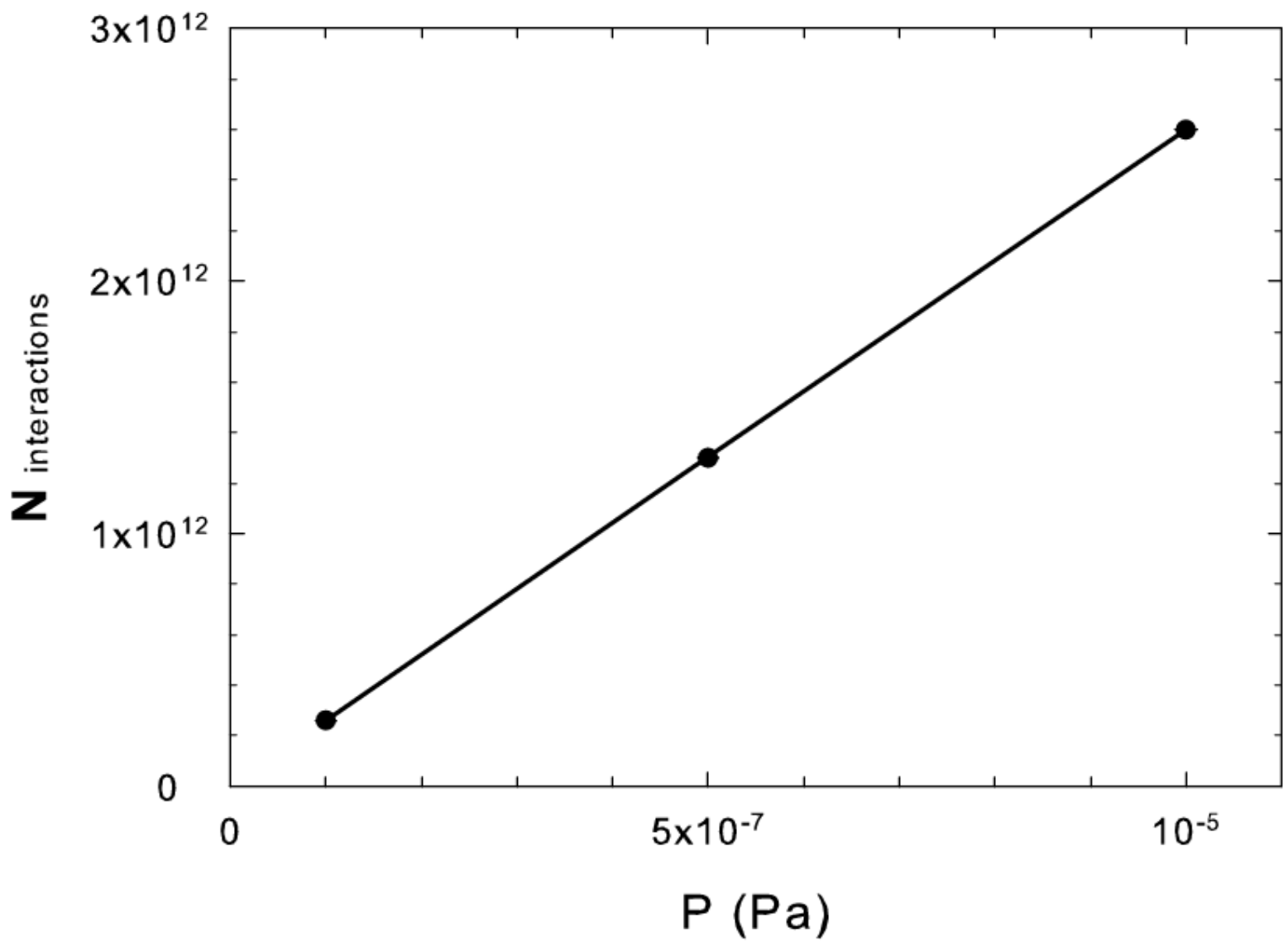

Fig.1 Dependence of number of beam particle interactions on residual gas pressure in accelerator chamber.

The above estimation must be considered as a rough idea. A realistic estimate must taken into account the effective pressure of the gas in the accelerator chamber (the pressure can be distributed not uniformly along the camber and in general case can vary with pressure), losses of particles on aperture of the chamber, chemical composition of the gas etc. These problems do not look as insurmountable ones. The measurements of electron scattering off infrared photons radiated by LEP accelerator chambers [1] demonstrated the great advantage of the approach.

In [2] there was made a proposal to study electron scattering off dark matter using evacuated tube as a target. With appropriate detectors such approach is more informative than the present one. But the single pass method severely limits the attainable cross section. Of course it can be cured if instead use a linear section of a synchrotron as a target.

In conclusion, the key point of the dark matter search is a reliable proof of it existence. The proposed method has the following advantages: 1) no special apparatus to record unknown particles and 2) possibility to measure rather low cross sections for relatively short time.

I would like to thank V. Parhomchuk for enlightening discussions.